\documentclass[doublecol]{epl2} 
\usepackage{amsmath}
\usepackage{amssymb}
\title{Playing against the fittest: A simple strategy that promotes the emergence of cooperation}
\shorttitle{} 

\author{Markus Brede\inst{1}}
\shortauthor{M. Brede}

\institute{                    
  \inst{1} CSIRO Marine and Atmospheric Research, CSIRO Centre for Complex System Science, F C Pye Laboratory - GPO Box 3023, Clunies Ross Street
Canberra ACT 2601, Australia\\

}


\abstract{Understanding the emergence and sustainability of cooperation is a fundamental problem in evolutionary biology and is frequently studied by the framework of evolutionary game theory. A very powerful mechanism to promote cooperation is network reciprocity, where the interaction patterns and opportunities for strategy spread of agents are constrained to limited sets of permanent interactions partners. Cooperation survives because it is possible for close-knit communities of cooperation to be shielded from invasion by defectors. Here we show that parameter ranges in which cooperation can survive are strongly expanded if game play on networks is skewed towards more frequent interactions with more successful neighbours. In particular, if agents exclusively select neighbors for game play that are more successful than themselves, cooperation can even dominate in situations in which it would die out if interaction neighbours were chosen without a bias or with a preference for less successful opponents. We demonstrate that the ``selecting fitter neighbours'' strategy is evolutionarily stable. Moreover, it will emerge as the dominant strategy out of an initially random population of agents.
}

\begin{document}

\maketitle

\section{Introduction}

A variety of social- and non-social organisms display altruism \cite{Crespi}. Altruism describes the ability to cooperate for joint benefit even at a cost to the individual. This scenario has frequently been studied by the prisoner's dilemma \cite{Axelrod}, in which two individuals simultaneously make a decision to cooperate (C) or defect (D). Cooperation gives a benefit $b$ to the recipient, but comes at a cost $c$ to the donor. Introducing the cost-benefit ratio $r=c/(b-c)$ and normalizing, a frequently used parametrization of the game is that mutual cooperation is rewarded by $R=1$ for both players. A cooperator interacting with a defector receives $S=-r$ while the defector obtains a payoff $T=1+r$. Mutual defection results in the same payoff $P=0$ for both opponents. For $0<r<1$ payoffs are ranked $T>R>P>S$ and $2R>T+S$, such that for one player independent of the other player's strategy, defection always results in a better outcome. The inferiority of the payoff for one-sided or mutual defection to mutual cooperation constitutes the dilemma setting. 

The evolution of cooperation is conveniently studied in the framework of evolutionary game theory \cite{Weibull}. One considers a population of agents whose strategies spread in the population according to their success in game play. Because defectors always earn larger payoffs than cooperators, cooperation cannot survive in well-mixed situations in which agents can interact with all other agents in the population. In this case defection is the evolutionarily stable strategy. As shown by studies of evolutionary games in space \cite{Nowak2,Lindgren1,Hauert1} and more recently on other types of complex networks \cite{Abramson,Santos4,Tomassini,Santos2,Santos3,Lieberman,Ohtsuki} the dominance of defection can be reduced if interaction patterns are restricted to a finite set of permanent opponents. Limited interaction patterns are conveniently described by spatial grids or complex networks such as scale-free or small-world networks \cite{Barabasi1,Watts1}. On a network, agents interact in playing the prisoner's dilemma game with their network neighbours. Strategy evolution is described by a process whereby agents can adopt the strategies of better performing neighbours. In such a setting cooperators can form tightly-knit clusters of cooperation that are shielded from the invasion of defectors. Nevertheless, even though cooperation is boosted by network reciprocity, for instance in the spatial game only limited numbers of cooperators can survive for a rather restricted range of dilemma strengths \cite{Hauert}.

Recent results on the evolution of cooperation on complex networks \cite{Abramson,Santos4,Tomassini,Santos2,Santos3,Lieberman,Ohtsuki,Masuda} have added substantial insight to the above statement. In particular, in situations in which some agents have the potential to earn substantially larger payoffs than others \cite{Santos2,Santos3,Perc,Helbing}, cooperation can be strongly enhanced, but is still not dominant in all possible game settings.

A mechanism to boost cooperation even on regular graphs similar to the heterogeneity in potential payoffs on heterogenous networks is via different degrees of effectiveness in spreading individual agents strategies' \cite{SzolSzab}, which can also be modelled as wealth accumulation of agents \cite{Helbing}. In a similar vein to our study and probably inspired by earlier work on dynamic rules for strategy selection \cite{Wu} also the aspiration of agents to perform better has been modelled in the context of the prisoner's dilemma game \cite{PercWang,WangPerc}. In their simulations, Wang and Perc assume that depending on aspiration levels agents preferentially select certain neighbours as potential targets to copy strategies. The procedure and additional heterogeneity in aspiration levels have been shown to further enhance cooperation. Different to these previous studies \cite{Wu,PercWang,WangPerc} the present Letter does not consider aspirations in the strategy adaptation process. Instead, we focus on the infuence of the strength of biases in the selection of game partners on the evolution of cooperation on fixed contact networks. In a series of controlled computer experiments we demonstate that a strong positive bias to preferentially interact with successful players strongly promotes cooperation.

\section{The Model}
Here we study a simple model of game partner selection on networks. As in the standard approach, contact networks define sets of potential interaction partners and candidates for strategy adoption. However, we assume that agents do not necessarily engage in game play with all of their network neighbours with the same likelihood, but may have a preference to select opponents for game play that are more or less successful than themselves. The success of an agent $x$ in playing the game is measured by its payoff $P_x$ from game interactions. In biological problems payoff translates into an individual's fitness which is often apparent in the invidual's physical appearance and might thus be easily judged by competitors.

Formally, we consider a model in which agents select game partners with a probability $P_\textrm{select}$ proportional to payoff differences to neighbours, i.e.
\begin{align}
\label{Pselect}
 P_{select} (z) \propto \frac{1} {1+\exp(w (P_z-P_x))}.
\end{align}
The parameter $w$ describes the preference of an agent to play against stronger ($w<0$) or weaker ($w>0$) neighbours. The case of no preference for game partner selection ($w=0$) corresponds to the setting in which each agent plays against all of its neighbours which is conventionally studied in the literature. But note that there is an important difference: even in unbiased probabilistic game partner selection not the whole neighbourhood of potential game partners might be explored and some agents can have more opportunities for game play than others. In the limits $w\to +\infty (-\infty)$ strategy choice simplifies to exclusively playing against weaker (stronger) opponents. In our setting, each agent probabilistically selects an opponent for game play as many times as it has neighbours. In each interaction, the focal and the selected player earn payoffs from playing the prisoner's dilemma game. Accumulated payoffs are then determined by counting each agent's payoffs, i.e. summing over payoffs from games in which the agent itself selected game partners and from games in which the agent was selected as a game partner.

In a second step strategies are updated after each round of game play. This is modelled in the conventional way. Every agent randomly selects one of its neighbours and adopts the strategy of the neighbour with a probability determined by payoff differences 
\begin{align}
P_{adopt}\propto 1/(1+\exp((P_z-P_x)/K))).
\end{align} 
The parameter $K$ gives the noise in the strategy selection process \cite{Hauert} and we set $K=0.1$ in the presented simulation experiments, but qualitatively similar results can be obtained for deterministic strategy adoption. After initializing the population with equal shares of randomly positioned cooperators and defectors, the iterated game play and strategy adaptation procedure is typically repeated for $T=10^5$ generations of system updates until a quasistationary state has been reached. Then averages are calculated over a further $T$ generations to determine average fractions of cooperators.

In the above procedure we chose a synchronous updating scheme because it gave faster convergence. Differences in synchronous and asynchronous updating can strongly influence results, particularly in purely deterministic simulations \cite{Huberman}. To address this problem we also made sure that the synchronous procedure does not bias the principal result. Qualitatively very similar outcomes are found for asynchronous updating.

Simulations were conducted for systems comprised of $N=10^5-10^6$ agents and results were averaged over between $10$ (random graphs) and $100$ (scale-free networks) realizations of the network topology when random networks are investigated. The fractions of average cooperators shown below represent averages over time (in the quasistationary state) and over different network realizations and initial conditions.
  
\section{Results}

\begin{figure*}[tbp]
 \begin{center}
\includegraphics[width=.95\textwidth]{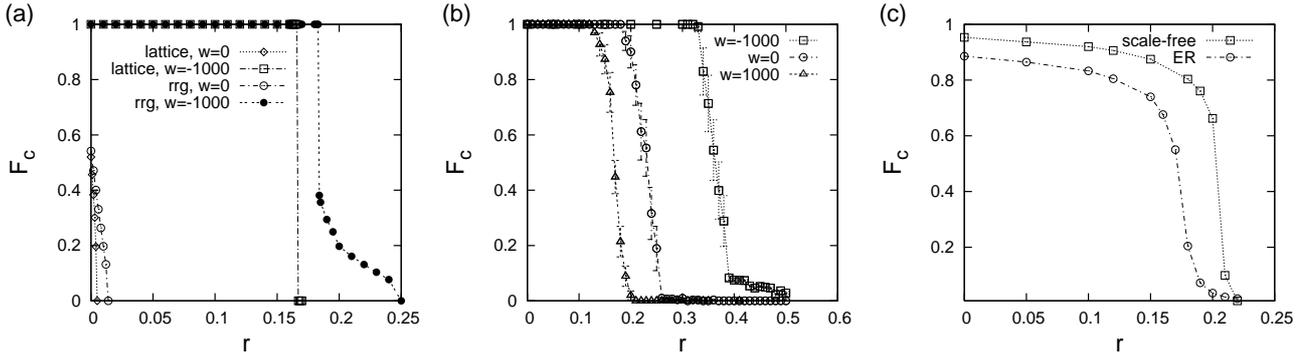}
 \caption{Dependence of the stationary number of cooperators $F_c$ on the cost-benefit ratio $r$. Compared are simulation results on: (a) lattices and regular random graphs (rrg) (b) scale-free networks and (c) scale-free networks and Erd\"os-R\'enyi random graphs (ER) with payoffs normalized by degree.  Notably, without game partner preference, no cooperation can be maintain beyond $r_c=0.005$ (lattices) or $r_c=0.013$ (regular random graph) and thresholds are $r_c=0$ in (a) and (c) for all classes of networks when preferences are for weaker opponents. The corresponding thresholds for strong game partner selection are $r_c=0.167$ and $r_c=0.25$ for lattices and regular random graphs.  Note also for strong game partner selection for $r<r_c$ complete cooperation prevails. The effect is also visible for scale-free networks, where cooperation is not completely subdued even if $w\to \infty$.}
\label{fig.0}
\end{center}
 \end{figure*}

Figure \ref{fig.0} summarizes some key results highlighting the influence of game partner selection on the evolution of cooperation on various classes of networks. On spatial grids we find that playing the game with strong preference for weaker opponents does not allow for the survival of any cooperators. Likewise, without preference in game partner selection for $r>r_c=0.005$ no cooperator can survive. These thresholds are lower than the respective thresholds for deterministic game play against all neighbours. The reason for this is that probabilistic game partner selection undermines cooperation in cases where a cooperator is surrounded by many defectors and only few cooperators. In some rounds of game play the focal cooperator may then lack support from the interactions with the adjacent cooperators and is thus easier overwhelmed by large numbers of defectors in its surrounding.

\begin{figure}[tbp]
 \begin{center}
\includegraphics[width=.45\textwidth]{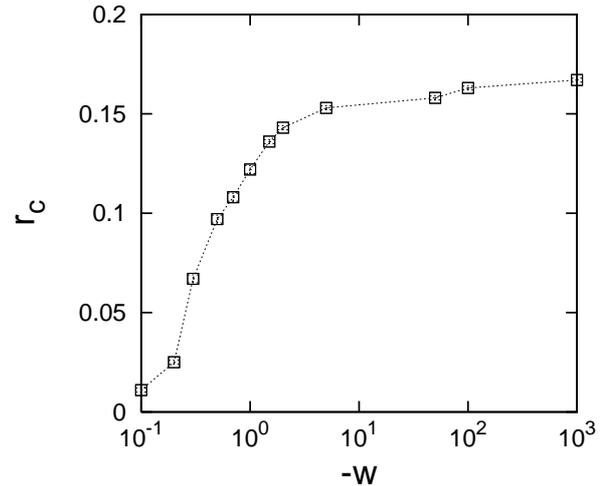} 
 \caption{Phase diagram for a lattice ($N=10000$, $K=0.1$). Increasing $-w$ allows cooperation to be maintained in an increasingly larger area of parameter space. Likewise, the transition becomes increasingly steeper and for large $-w$ is is virtually steplike between full and no cooperation.}
\label{fig.1}
\end{center}
 \end{figure}

This diminished ability of cooperators to survive with neutral selection is in stark contrast to the case of strong preference for the selection of stronger opponents. In the latter case cooperation dominates strongly up to cost benefit ratios of $r_c=0.16$. A systematic investigation of the dependence of the critical point above which cooperation cannot surive on the preference parameter $w$ is illustrated in Fig. \ref{fig.1}. The systematic shift of the critical point $r_c$ at which cooperators go extinct with increasing bias $w$ to select stronger game partners further supports our assertion.

Qualitatively similar results also hold for other classes of contact networks, e.g. for regular graphs or scale-free networks. This is independent of whether hub nodes can achieve larger payoffs than low degree nodes (which is known to be an effect that strongly promotes cooperation \cite{Santos2,Santos3}) or whether payoffs are normalized by degree (which diminishes the advantage of heterogenous networks \cite{Santos4}). However, it is important to stress that the boost for cooperators is limited to sparse networks and won't be found in a well-mixed population in which no shielded pockets of cooperators can form.

\begin{figure}[tbp]
 \begin{center}
\includegraphics[width=.45\textwidth]{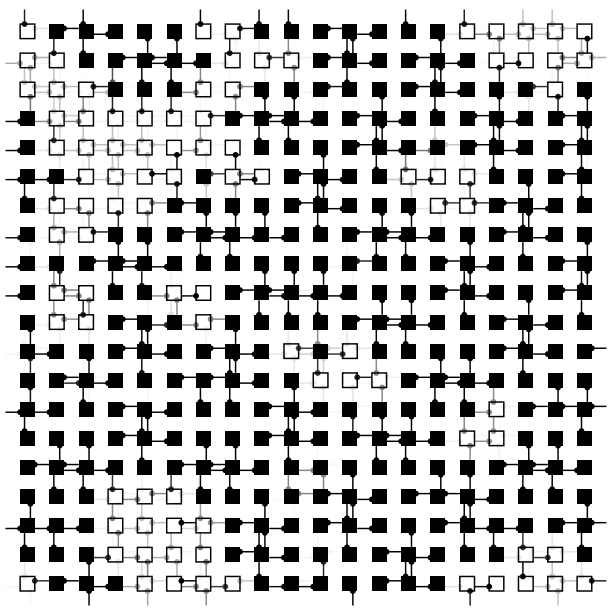} 
 \caption{Illustration of the flow of selection probabilities from game partner selection. The figure represents a $20 \times 20$ lattice with periodic boundary conditions, white boxes symbolize defectors and black boxes cooperators. Lines give the likelihood to select the linked neighbours for game play: white lines correspond to selection with probability zero and black lines game partner selection with probability one. Black dots indicate the respective reference nodes whose selection preferences the lines indicate.  }
\label{fig.n}
\end{center}
 \end{figure}

At the first glance these results may appear counterintuitive. One would assume that defectors can achieve larger payoffs and hence conclude that game partner selection for strong opponents might favour game play with defector nodes. This would then imply an additional advantage for defectors which would reinforce their dominance. This argument certainly holds in well-mixed populations, but becomes invalid for ordered arrangements of cooperators and defectors. Once clusters of cooperators have formed, selection for strong game partners can effectively shield clusters of cooperators from the invasion of defectors. This is so because cooperators within the cluster attract interactions with cooperators at the cluster's boundary. In this way payoffs of cooperator nodes inside the cluster and at its fringe are enhanced and interactions with defectors are avoided. Similarly, for defectors surrounding clusters of cooperators the effect limits opportunities for exploitation, which opens up the way for clusters of cooperators to expand. This is illustrated in Fig. \ref{fig.n}, which gives a snapshot view of the flow of game partner selection preferences. 

In a similar vein it is easy to understand why the ``selecting weaker opponents'' strategy undermines cooperation. Whereas it helps cooperators to seek out fellow cooperators for interactions in an initially well-mixed population it also allows defectors to avoid interactions with other defectors and thus helps them find cooperators to exploit. The latter effect together with enhanced exploitation of fringe cooperators in ordered populations explains the ineffectiveness of the strategy.

Another argument why ``playing stronger opponents'' supports cooperation is related to the boost cooperators experience from heterogeneous settings \cite{Santos2,Santos3,Perc,Helbing}. As the strategy spread is a process of replacement of less successful strategies by more successful ones, ``playing stronger opponents'' effectively modifies network topology and enhances game play between pairs of agents between which frequent strategy copying takes place. In this way, a successful defector undermines its own success quickly by first attracting interactions that further enhance its success, but then quickly spreading its strategy to its neighbours who defect in interactions with it. Contrariwise, the process enhances support for a successful cooperator.

\begin{figure*}[tbp]
 \begin{center}
\includegraphics[width=.95\textwidth]{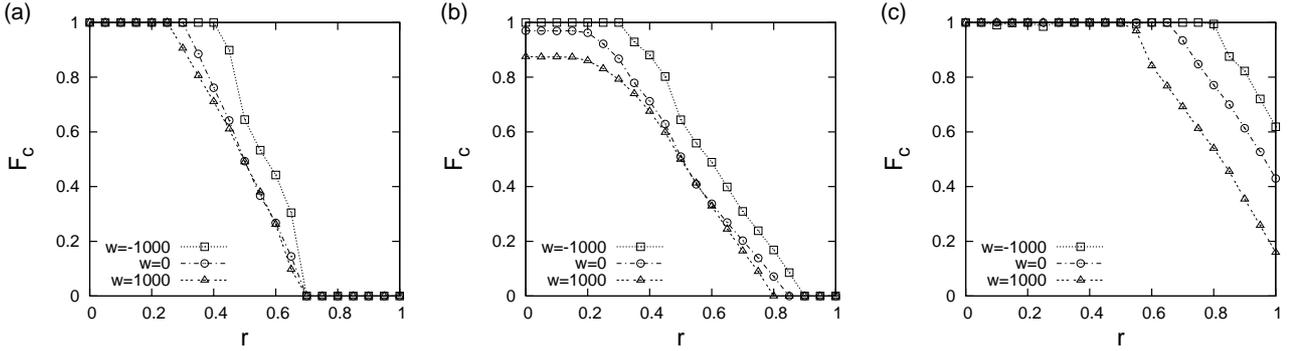} 
 \caption{Evolution of cooperation in the snowdrift game. Comparison of rules with $w=1000$ (up triangles), $w=0$ (circles), and $w=-1000$ (boxes) for (a) regular grids, (b) regular random networks and (c) scale-free networks.}
\label{fig.s1}
\end{center}
 \end{figure*}

For some biological applications milder dilemma settings than the prisoner's dilemma are of interest (see \cite{Hauert1} and references therein). A typical scenario is described by the snow-drift game, which is characterized by a ranking of payoffs $T>R>S>P$. Due to the changes of $S$ and $P$ in payoff rankings some proportion of cooperators can survive even in well-mixed populations. The snow-drift game is also of interest, since network reciprocity does not necessarily enhance cooperation compared to the well-mixed case \cite{Hauert1}. We explored the parameter range of snow-drift dilemma games on various classes of networks. For all investigated situations, spatial, small-world and scale-free networks we find that game partner selection with a bias towards strong opponents substantially enhances the number of surviving cooperators, see figure \ref{fig.s1}.

\section{Evolution of biases}
Let us consider an experiment in which we start with a population in which no bias for game partner selection exists. Initially every agent is either a defector or a cooperator with equal probability and might also have some individual bias for game partner preference. This can be modelled by an individual preference parameter $w_i$ that is initialized from a uniform distribution over the intervall $[-1,1]$. Assume that the game partner selection preference is hereditary (similar to the evolution of strategies). Would a systematic game partner preference in the population evolve? And if so, in the face of invasions of new agents with randomly selected game partner preference and strategy profile, would it be stable?

\begin{figure}[tbp]
 \begin{center}
\includegraphics[width=.49\textwidth]{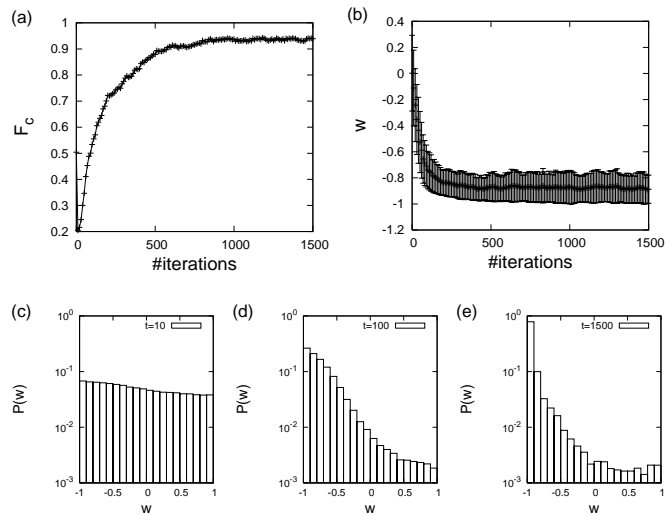} 
 \caption{The evolution of cooperation in the prisoner's dilemma game from initially random game partner preference. (a) Evolution of the frequency of cooperators. (b) Evolution of the average preference of the population, error bars indicate the variance and (c-e) Histograms for the distribution of game partner selection preferences in the population at differentstages of the evolution, i.e. after $t=10$, $100$ and $1500$ steps. }
\label{fig.2}
\end{center}
 \end{figure}

Following similar approaches like \cite{PercWang,Szolnoki} we systematically addressed these questions in a number of computer experiments.  To study the evolution of game opponent selection preferences we initialize a population of agents randomly, attributing every agent with a strategy for game play (cooperation or defection) and an individual game partner selection preference $w_i \in [-1,1]$. Game play and strategy updating were modelled as above with three important differences. First, agents select game partners as determined by their individual preference $w_i$. Second, in the strategy adaptation process agents not only adopt their neighbours' strategies for game play but also their respective game partner selection preferences. Third, we introduced a small chance for mutations. After strategy updating with small probability $q=10^{-3}$ an agent will assume a randomly selected combination of game play strategy and game partner selection preference.

Figure \ref{fig.2} illustrates a typical outcome and gives several snapshots during the evolution. After half of the population was seeded with cooperators their frequency drops quickly in the initial timesteps of the simulation. The drop occurs because of the initially random mixing in which no pockets of cooperators have evolved yet. During this stage of the evolution game preferences are uniform and there is yet no bias for selecting stronger opponents. However, already after $100$ sweeps of updates cooperators have recovered to occupy half of the population and the distribution of $w$'s exhibits a strong bias towards negative values. The bias becomes stronger as cooperation thrives more and more and the distribution of $w$'s only retains a tail towards positive values of $w$ since we allowed for invasions of agents with randomly selected strategy profiles with a small rate $q=10^{-3}$ per site update. These random invasions also prevent the system from reaching full cooperation. 

A closer investigation reveals the mechanism by which game partner preferences come to dominate from random strategy distributions. Clearly, by the above arguments cooperators at the fringe of  pockets of cooperation fare better if they select for strong game opponents. Cooperators with differing biases will be likely to either  be invaded by defectors or to copy the strategy of other cooperators with negative bias in $w$ since the latter tend to fare better.  Thus cooperation is reinforced and can spread. Cooperators within pockets of cooperation can initially sustain a neutral or positive game partner selection bias, but will be subject to the same selective pressure if they become fringe nodes when mutations towards defection occur in their neighbourhood.

These simulations emphasise two important points. First, preferences in game partner selection will evolve from random initial configurations under very mild assumptions. Second, once evolved, they are evolutionarily stable and even robust in the presence of substantial noise, either in the strategy updating process or from random invasions of defectors.

\section{Conclusions}

In summary, we have demonstrated that a simple rule for game partner selection can significantly enhance the support for cooperation on networks. The simple and biologically plausible rule to play preferentially against better-off opponents concentrates game play interactions within clusters of cooperators and thus strengthens them against invasions by defectors. This simple rule is found to be valid on all investigated classes of networks and boosts cooperation independent of whether payoffs are normalized by agent's individual numbers of interactions per round or not.

In comparison to the case of deterministic game play against all neighbours the rule not only significantly shifts the critical parameter $r_c$ that demarcates the parameter region where cooperation can persist, but also allows cooperation to dominate below the threshold. The effect is particularly pronounced for strong game partner preferences on lattices, on which cooperators can take over the entire population for $r<r_c$.

As a further point we also investigated the emergence and evolution of game partner preferences. We demonstrated that the `playing against fitter opponents' rule can emerge dominant even from a randomly composed starting population consisting of equal shares of agents with positive and negative preferences. These results point to the potential importance of biases in opponent selection and complement recent research on aspiration levels of agent populations. Preliminary results point out that via a suitable combination of game partner preferences and aspiration biases any prisoner's dilemma situation can be resolved such that cooperation takes over the entire population. A more detailed investigation of the latter situation is left for future work.

\section{Acknowledgements}
This research was undertaken on the NCI National Facility in Canberra, Australia, which is supported by the Australian Commonwealth Government. I thank A. Schapper for helpful comments.

\end{document}